# Benchmarking database performance for genomic data

Matloob Khushi

Bioinformatics Unit, Children's Medical Research Institute, Westmead, NSW Australia. Centre for Cancer Research, Westmead Millennium Institute; Sydney Medical School, Westmead, University of Sydney, Australia

Email: mkhushi@uni.sydney.edu.au

**Abstract**

Genomic regions represent features such as gene annotations, transcription factor binding sites and epigenetic modifications. Performing various genomic operations such as identifying overlapping/non-overlapping regions or nearest gene annotations are common research needs. The data can be saved in a database system for easy management, however, there is no comprehensive database built-in algorithm at present to identify overlapping regions. Therefore I have developed a region-mapping (RegMap) SQL-based algorithm to perform genomic operations and have benchmarked the performance of different databases. Benchmarking identified that PostgreSQL extracts overlapping regions much faster than MySQL. Insertion and data uploads in PostgreSQL were also better, although general searching capability of both databases was almost equivalent. In addition, using the algorithm pair-wise, overlaps of >1000 datasets of transcription factor binding sites and histone marks, collected from previous publications, were reported and it was found that HNF4G significantly co-locates with cohesin subunit STAG1 (SA1).

**Introduction**

The recent revolution in whole genome census approaches has seen an exponential increase in available data sets describing genomic features, such as transcription factor binding sites and histone modifications. Curation of such data and identifying relationships, such as overlaps in genomic features and closest gene annotation, are fundamental tasks in this research [Meyer et al., 2012]. The files containing such genomic data usually have chromosomal location (chromosome, start and end) information in them. A number of tools such as Galaxy [Goecks et al., 2010], BedTools, GenomicTools [Tsirigos et al., 2012] and BEDOPS Tools [Neph et al., 2012] have been developed to find overlapping/non-overlapping nearby regions [Neph et al., 2012; Quinlan and Hall, 2010; Zammataro et al., 2014]. The relationships among the files are usually manually managed. With exponential growth in available genomic information, managing manual relationship and curation of such files are becoming more cumbersome day by day. These relationships and curation can be better managed using a relational database such as Microsoft SQL Server, Oracle, MySQL or PostgreSQL; however, there is no dedicated published algorithm available that is natively built into a database system to operate on the genomic features. I have therefore developed a novel algorithm Region Mapping (RegMap) that operates on genomic locations natively in the database and have benchmarked the performance of two major open-source free databases PostgreSQL and MySQL. I have also compared the RegMap performance against database built-in spatial functions which provide very limited functionality.

**Methods**

Two genomic regions (genomic intervals) are said to intersect or overlap if both intervals share at least one base pair in common on a chromosome. Chromosomes were saved as character data-type and start and end coordinates as integer data-type for the RegMap algorithm. To compare performance with database built-in spatial functions the coordinates were saved as linear spatial data-type. Genomic regions were saved in the *Regions* table and were linked with the *RegionDesc* table where annotation of the regions was saved, thus simulating a production usage. Each region in the *Regions* table was automatically assigned a unique database id (Primary Key). The start coordinates of the genomic regions were indexed from 0, according to UCSC recommendations (http://genome.ucsc.edu/) to speed up calculations, therefore region length was calculated by subtracting the start from the end coordinate.

RegMap generates all the required objects in a working database. The algorithm was developed in native SQL (Structured Query Language) and is therefore compatible with all SQL databases.

A total of 1005 datasets of transcription factor binding sites and histone marks from previous publications on human and mouse assemblies were collected, including hg19, hg18, mm9 or mm8. This 'Knowledge Base' was used to perform search benchmarking.

All testing and benchmarking were performed on PostgreSQL 9.0 and MySQL Community Server 5.6.15 GPL (x86_64) installed on a personal computer of 4 core 2.4 GHtz processor with 8GB memory. MySQL Server supports a number of storage engines, however I have benchmarked performance for two widely used InnoDB and MyISAM storage engines [Padilla and Hawkins, 2011; Sheldon and Moes, 2005]. The results of 100 simulations were averaged for all operations. RegMap code was run in MySQL Workbench 6.1 for MySQL server and in pgAdmin III 1.81 for PostgreSQL benchmarking maintaining the default settings of each database. The default random region size was set to 500, however, this setting can be changed in the script.

**Results**

**Development of the RegMap algorithm**

RegMap finds overlapping or non-overlapping regions by calculating the number of bases common or away between two regions. Therefore, a variable 'bp overlap' was devised which was calculated positive (shown as shaded regions in the Figure 1) by counting the number of base pairs in common between two regions or calculated negative when away from the ends of the two regions. An illustration of the algorithm can be found in Figure 1. There are three possibilities:

i) **One region is within, or completely overlaps, the other.** In this case the bp overlap is simply a positive number reported by calculating the length of the smaller region that lies within the second region. If the two regions completely overlap each other then the length of either region can be reported as the bp overlap. For example, where region A lies within region B (Figure 1-i-a), this can be identified computationally by checking if A.End is less than or equal to B.End and if A.Start is greater than or equal to B.Start. The region length of A can be calculated by the SQL pseudocode extract given below:

```
WHEN A.End <= B.END AND A.Start >= B.Start
    THEN (A.End - A.Start)
```

Conversely, when region B lies within region A (Figure 1-i-b) or completely overlaps, this can be confirmed by checking whether B.End is less than or equal to A.End and if B.Start is greater than or equal to A.Start. The region length of B can then be calculated:

```
WHEN B.End <= A.END AND B.Start >= A.Start
    THEN (B.End - B.Start)
```

ii) **Region A is located on the left side of region B**. In this possibility the two regions may share bases in common (Figure 1-ii-a) or can be completely away from each other (Figure 1-ii-b). Computationally this is verified by checking whether A.End is less than or equal to B.End and A.Start is less than or equal to B.Start. The bp overlap is calculated by subtracting B.Start from A.End:

```
WHEN A.End <= B.End AND A.Start <= B.Start
        THEN (A.End - B.Start)
```

Using the above calculation, for the first situation (Figure 1-ii-a) the bp overlap will be reported as a positive integer. For the second situation (Figure 1- ii-b) when no common bases exist between the two regions, the bp overlap is returned as a negative number. This is because the B.Start coordinate is greater than A.End therefore `(A.End - B.Start)` will be a negative number.

iii) **Region A is located on the right side of region B**. As above, the two regions could overlap or can be apart without intersecting each other. This is verified by checking whether A.End is greater than or equal to B.End and if A.Start is greater than or equal to B.Start. The bp overlap is calculated by subtracting A.Start from B.End:

```
WHEN A.End >= B.End AND A.Start >= B.Start
        THEN (B.End - A.Start)
```

Similar to the above case the overlapping regions (Figure 1-iii-a) will have positive bp overlap and non-overlapping regions will have negative bp overlap (Figure 1 iii-b).

I also calculated the distance between the centres of two regions, referred to as 'centre distance'. Sometimes a small base pair overlap of very long regions does not make any biological sense so overlap analysis should be limited to a certain distance from the centre of two regions. This is also useful in extracting regions that do not overlap but are very close to each other. To calculate the distance, the centres of the two regions are determined and then the absolute (positive) distance between the centres is calculated by the following SQL code extract.

```
abs ( (A.End + A.Start)/2 - (B.End + B.Start)/2 )
```

The full code for the algorithm has been provided in Supplementary File 1.

**Benchmarking for insertion of genomic regions**

RegMap script randomly generates region data between the specified lower and upper range which is temporarily saved in memory and then saved in the database in a single transaction. This technique was faster for both databases, since each time an *insert* statement was executed against the database there were transaction overheads. Therefore, generating and saving regions one by one was much slower than saving all the data at once.

I tested the performance by generating 5K, 10K, 20K, 40K and 80K regions and identified that PostgreSQL's generation of random regions and insertion was much faster than MySql in both InnoDB and MyISAM storage engines. MySQL's insertion of regions was dramatically slower and the time taken was almost double by doubling the number of regions inserted (Figure 2). MySQL-InnoDB performed slightly better than MySQL-MyISAM, therefore, in Figure 2 performance of MySQL-InnoDB is reported. PostgreSQL (RegMap) generated and saved 5K random regions in 1 second as compared to 219 second in MySQL-InnoDB and 237 second in MySQL-MyISAM, indicating that MySQL was ~220 times slower. This difference dramatically increases for a much larger number of regions. For generating 80K random genomic regions PostgreSQL took 4 seconds as compared to 3,596 sec in MySQL-InnoDB and 3,680 second in MySQL-MyISAM.

In addition, the write performance was tested by importing the 1005 files consisting of 23,827,431 real genomic regions, collected from previous studies, into both databases using bulk import statements of the databases. PostgreSQL *COPY* statement while MySQL *LOAD DATA INFILE* statement was used for this purpose. I performed the import of each file in three steps: i) data was imported into a staging table, ii) data was copied across the production table while assigning a unique id, and iii) the staging table was emptied. This procedure was adopted because in reality the imported data usually needs to be processed and assigned a unique identity in order to link to other tables. PostgreSQL performed >5 times better than MySQL, PostgreSQL took ~445 seconds compared to ~2,940 seconds in MySQL-InnoDB and 2,460 second in MySQL-MyISAM. The actual import script is also provided in the Supplementary File 1.

Data upload performance is critical for bioinformatics servers where many users insert a large amount of data at once. Therefore this benchmark identified that PostgreSQL inserts and imports data much faster than MySQL.

**Benchmarking for identification of overlapping regions**

I further investigated the performance of reporting intersecting or overlapping regions using RegMap and using the database built-in functions in each database. The two databases have built-in functions that can be used to identify intersecting lines. Since these built-in functions are usually used in geographical (spatial) mapping software I subsequently refer to the built-in functions as Geo functions.

PostgreSQL's performance was again outstanding in finding overlapping genomic regions compared to MySQL (Figure 3). RegMap in PostgreSQL took 134 seconds to report intersecting regions for two datasets of 80K regions each, and 257 seconds using PostgreqSQL-Geo function. MySQL performance was tested using InnoDB and MyISAM storage engines. MySQL-MyISAM performed poorly compared to InnoDB, however, both engines demonstrated inferior performance as compared to PostgreSQL. For example, when two datasets of 80K regions were tested for overlaps using RegMAP, MySQL-InnoDB took 1,119 seconds, and MySQL-MyISAM took 1,150 seconds. Therefore, for simplicity reasons, I presented the data for MySQL-InnoDB in Figure 3.

Applying various indexes on these regions did not improve performance in PostgreSQL while it had a negative effect in MySQL for both engines (InnoDB & MyISAM). I performed these tests on different computers and obtained slightly different timings, however, the overall outcome remained the same which was that PostgreSQL performance in identifying overlapping regions was much better than MySQL.

Since RegMap identifies overlapping regions by calculating 'bp overlap' for each region, I finally concluded that queries that require extensive calculation of mathematical operations perform much better in PostgreSQL.

**Searching and retrieving regions**

PostgreSQL was slightly better at performing a search of genomic regions than MySQL. The knowledge base of ~24 million genomic regions was searched for erroneous regions with a start coordinate less than 0 or an end coordinate less than start. PostgreSQL identified 10 erroneous regions in ~5 seconds while MySQL-InnoDB found the same erroneous regions in ~21 seconds and MySQL-MyISAM in 6 seconds. Implementing various types of indexes on chromosome start and end fields did not improve performance for this query. However, searching for specific regions within a certain distance of a gene was instant in all databases. For example, searching regions within 100,000 upstream/downstream of the transcription start site of MYC gene (chr8:128748314) returned results in 3-5 seconds in all databases which was further reduced to ~1 second by implementing an index. Therefore I concluded that the general searching capability of PostgreSQL and MySQL is similar. The Queries are provided in the Supplementary File 1.

**Advantages of RegMap over Geo functions**

The RegMap algorithm not only outperformed the Geo functions, in addition, it provides extended functionality that the Geo functions do not provide. For example, Geo functions only return a Boolean (true/false) value if the queried regions intersect or not. On the other hand, RegMap reports the number of bases common in the two regions or away from each other. Therefore it is easy to limit results based on the minimum number of bases that must overlap. It also provides the ability to restrict results based on the distance from the centre of regions; this is useful in returning regions that do not share common bases, but are present in close proximity within a specified distance. For example the SQL query *select * from vwregions where bpoverlap<1 and centredistance<1000;* will return regions that do not overlap however their centres are within a distance of 1000 bases. This type of analysis is important in identifying partner factors that bind on DNA in close proximity to each other without overlapping.

**Mining Knowledge Base**

The Supplementary File 2 contains overlapping results of >1000 datasets of transcription factor binding sites and histone marks with information about cell line, treatment condition (if there is any), total number of regions, the number of overlapping regions and its percentage found against other datasets (Supplementary File 2). The results are spread across four spreadsheets based on the aligned reference genome i.e hg19, hg18, mm9 or mm8. Researchers can easily filter records based on restricting values in each field and then sorting on 'Percentage Overlaps' to find out the most or least interacting dataset. There are links provided to the raw data and to the publications.

This useful knowledge base can help in developing new hypotheses that can further be tested and analysed in the wet lab. For example, using the knowledge base a novel cis-regulatory interaction between estrogen receptor alpha (ERα) and progestin receptor (PR) was identified [Khushi et al., 2014a]. I observed that among all factors in hg18 assembly SA1, Rad21 and CTCF binding locations were comparatively conserved in MCF7 (breast adenocarcinoma cell line) when compared to their binding locations in H1 hESC (human embryonic stem cell line).These factors targeted similar genomic locations in the two distinct cell-lines, despite previous reports describing the binding pattern of various factors to be cell specific [Consortium, 2012; ENCODE, 2012; Wang et al., 2012].

In the HepG2 (liver hepatocellular) cell line, I identified that HNF4G preferentially binds (6633/6839, 97%) to H3K4Me1 (enhancer) regions, and the majority of HNF4G binding sites (4244/6839, ~62%) were also found overlapping with STAG1 binding sites (83080 regions). The statistical significance of overlapping of HNF4G is further analysed in BiSA [Khushi et al., 2014b] which revealed a statistically significant overlap correlation value of 0.65. As previously described, the overlap correlation value greater than 0.5 shows a statistical significant overlap of a query factor [Khushi et al., 2014b]. HNF4G is an orphan nuclear receptor whose ligand and function has not been fully understood, however recent studies have shown HNF4G overexpression to induce growth of cancer tissue [Okegawa et al., 2013; Yang et al., 2014]. On the other hand, STAG1 (Stromal Antigen 1), also known as SA1, is one of the four subunits of the cohesin complex [Losada, 2014]. Cohesin has important roles in transcription regulation, DNA repair, chromosome condensation, homolog pairing, etc. [Losada, 2014; Mehta et al., 2013]. Therefore, statistical significant overlap of HNF4G with STAG1 indicates an important underlying biology which could be further explored in laboratory.

**Discussion**

Various databases are heavily used in biomedical and cellular biochemistry, therefore researchers would benefit from knowing which database product performs better for a specific type of data. Benchmarking software products also helps vendors to improve their products. Various other benchmarks for database systems exist and it is acknowledged that development and adoption of benchmarks advances research in a research area [Aniba et al., 2010; Arslan and Yilmazel, 2008; Bose et al., 2009; Ray et al., 2011; Sim et al., 2003; Venema et al., 2013]. For example, Ray et al. [Ray et al., 2011] benchmarked databases for spatial data, whereas, Xu et al. [Xu and G¨uting, 2012] benchmarked databases for moving objects data. However no benchmarking effort exists on database performance for genomic region operations. Therefore RegMap, being natively written in SQL and adaptable for any SQL-based database, will advance research in this field and will provide a baseline mark for future algorithms.

Many bioinformatics analyses produce a large number of variant files. Usually detailed information about factor, cell-line, condition, peak-calling or analysis parameters used are recorded as part of file names or kept separate which makes it difficult to manage such information for a large scale study. Databases provide a more effective way of managing curation, annotation, sorting and relationships among data. RegMap, being a SQL-based algorithm, can be integrated into any language as most languages provide API (application programming interface) to connect to SQL-based databases. SQL's simple syntax is also easy for biologists to learn. There are a number of tools that are in use by the research community to operate on genomic regions, for example BEDTools [Quinlan and Hall, 2010], Pybedtools [Dale et al., 2011], GenomicTools [Tsirigos et al., 2012], and BEDOPS Tools [Neph et al., 2012]. All of these tools are designed to operate on files and integration of these tools in other languages is usually difficult. Tabix [Li, 2011] is another efficient tool that is usually used to extract specific regions from large files. The database capability of searching specific genomic regions was equivalent to Tabix. Both databases, when searching a table with ~24 million real genomic regions, returned results in ~1 second for regions that were within 100K of transcription start site of MYC gene. There are only a few tools which provide an easy interface to other languages, such as Pybedtools which provides Python interface, and GROK and GenomicTools which provide C++ API (application programming interface) to C/C++ programmers. There are a few GUI (Graphical User Interface) tools such as Cisgenome [Ji et al., 2011], Galaxy [Goecks et al., 2010], and the UCSC table browser [Karolchik et al., 2004] which provide very basic genomic operation analysis options. However, there is no algorithm available that performs genomic region operations natively in a relational database system. Therefore direct comparison of the performance between RegMap algorithm and other tools that work on files is not appropriate.

RegMap, being an SQL based algorithm, is easy to apply on an unlimited number of datasets. The algorithm was applied to ~1000 datasets of transcription factor binding sites and epigenetic marks and and an easy navigate-able Excel file was generated.

This data can be used to develop new hypotheses such as identification of novel biochemical partners of a factor or factor's binding influenced by a histone mark. Once an interesting interaction is found, actual genomic locations could be studied using tools such as BiSA [Khushi et al., 2014b]. Using the knowledge base a novel interaction between HNF4G and cohesion subunit STAG1/SA1 was identified. The majority of HNF4G binding sites overlapped STAG1 and this overlap was statistically significant suggesting an important biochemical partnership on a specific subset of genomic regions.

The performance of proprietary databases such as Oracle or Microsoft SQL Server was not reported because of their licensing restrictions, however, using the algorithm researchers/institutions can benchmark the suitability of either product for their own use.

I acknowledge that in other computational infrastructure database performance could vary, however, I have done rigorous testing on different machines to conclude that similar relative results would be obtained. The results also revealed that there is a great deal of room present to improve the database built-in functions that are used to find intersecting geometrical shapes. In other fields such as geo mapping application it is usually not required to find the thousands of intersecting features. However genomic studies deal with a large amount of data. With increased use of databases in genomic applications, there is a need for database functions to be improved for genomic operations. Therefore it is proposed here that 'genomic region' data-type in all databases should be implemented.

In summary, RegMap is an open source database-driven algorithm used to find overlapping/non-overlapping regions, and results can be limited by the number of bases in common or maximum distance between the centres of two sets. Using the algorithm I benchmarked performance of two widely used open source databases, PostgreSQL and MySQL. The benchmark revealed that PostgreSQL performs much better in identifying overlapping genomic regions. Data upload/import function of PostgreSQL was also better than MySQL. Data upload performance is critical for bioinformatics facility servers where many users insert a large amount of data simultaneously. Using the algorithm the overlapping of >1000 datasets of transcription factor binding sites and histone modifications were calculated and identified that HNF4G binding significantly overlaps with cohesion subunit STAG1/SA1 binding on DNA.

**Acknowledgements**

The author was supported by an Australian Postgraduate Award (APA) and Westmead Medical Research Foundation (WMRF) Top-Up scholarship. The author would like to deeply thank Professor Christine L. Clarke and Dr. J. Dinny Graham for their supervision during PhD candidature and to Dr. S. Cunningham for correcting some grammatical errors.

Sheldon R, Moes G. 2005. Beginning MySQL. Wiley.

Sim SE, Easterbrook S, Holt RC. 2003. Using benchmarking to advance research: A challenge to software engineeringeditor^editors. Proceedings of the 25th International Conference on Software Engineering. IEEE Computer Society, p 74-83.

Tsirigos A, Haiminen N, Bilal E, Utro F. 2012. GenomicTools: a computational platform for developing high-throughput analytics in genomics. Bioinformatics 28:282-283.

Venema V, Mestre O, Aguilar E, Auer I, Guijarro J, Domonkos P, Vertacnik G, Szentimrey T, Stepanek P, Zahradnicek P. 2013. Benchmarking homogenization algorithms for monthly dataeditor^editors. AIP Conference Proceedings. p 1060.

Wang J, Zhuang J, Iyer S, Lin X, Whitfield TW, Greven MC, Pierce BG, Dong X, Kundaje A, Cheng Y, Rando OJ, Birney E, Myers RM, Noble WS, Snyder M, Weng Z. 2012. Sequence features and chromatin structure around the genomic regions bound by 119 human transcription factors. Genome Research 22:1798-1812.

Xu J, G¨uting RH. 2012. GMOBench: A Benchmark for Generic Moving Objects. Informatik-Report 362, Fernuniversität Hagen.

Yang CS, Chang KY, Rana TM. 2014. Genome-wide functional analysis reveals factors needed at the transition steps of induced reprogramming. Cell Rep 8:327-37.

Zammataro L, DeMolfetta R, Bucci G, Ceol A, Muller H. 2014. AnnotateGenomicRegions: a web application. BMC Bioinformatics 15:S8.

- 16 -

# Figures

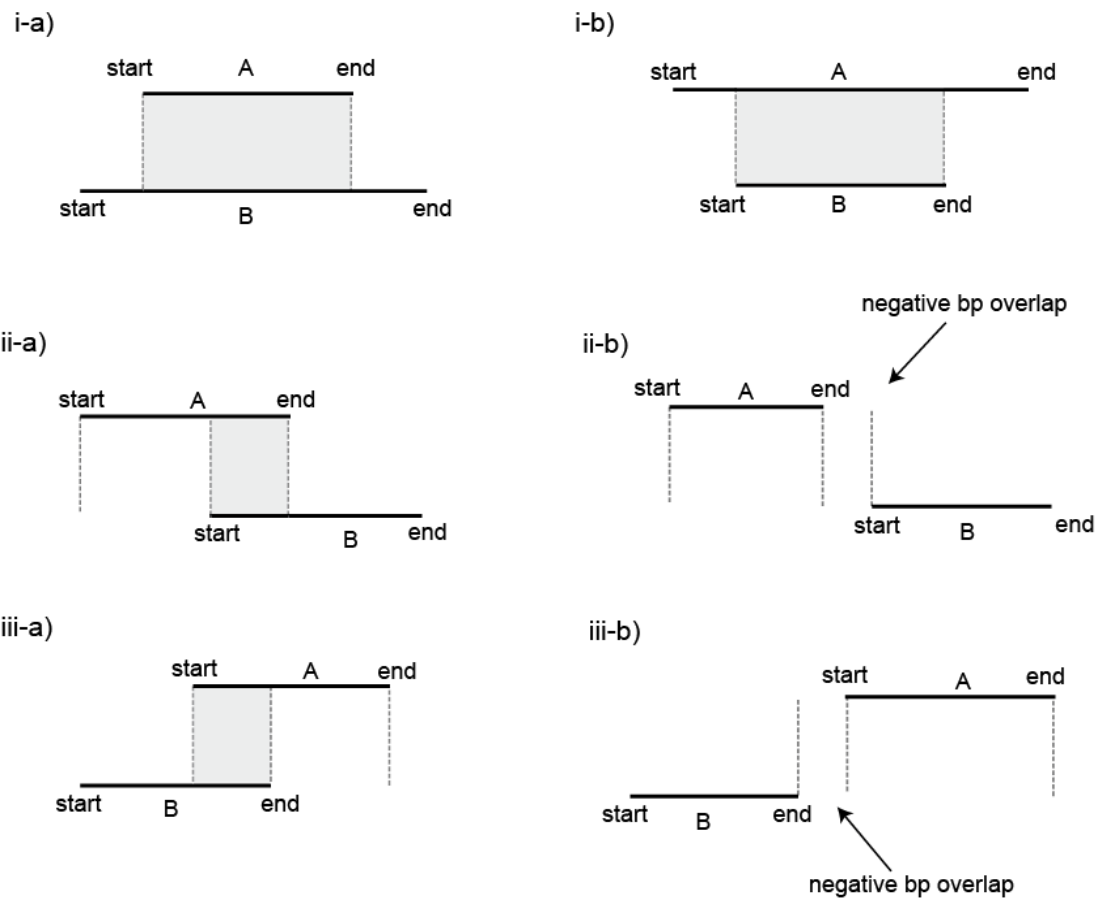

**Figure 1: Various possible relative positions of the two genomic regions.**

i) One genomic region is completely within the other. ii) The overlapping or non-overlapping region A is on the left side of the region B. iii) The overlapping or non-overlapping region A is on the right side of the region B.

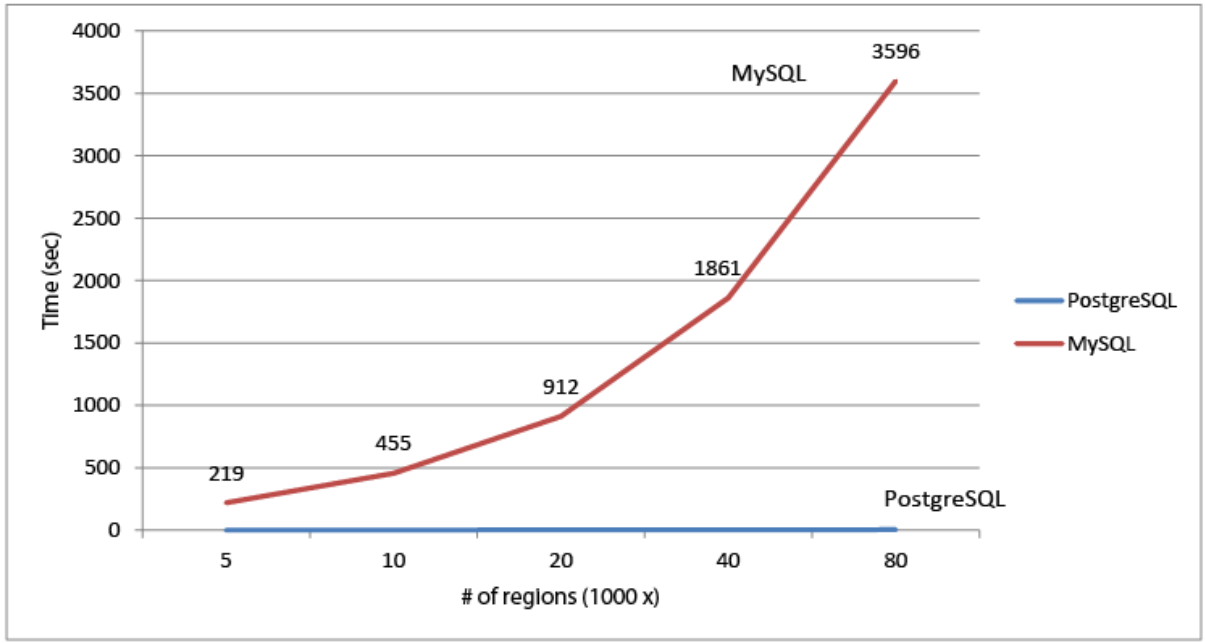

**Figure 2: Comparison of region insertion performance.** For simplicity, MySQL times shown are for InnoDB storage engine as MyISAM did not perform as well.

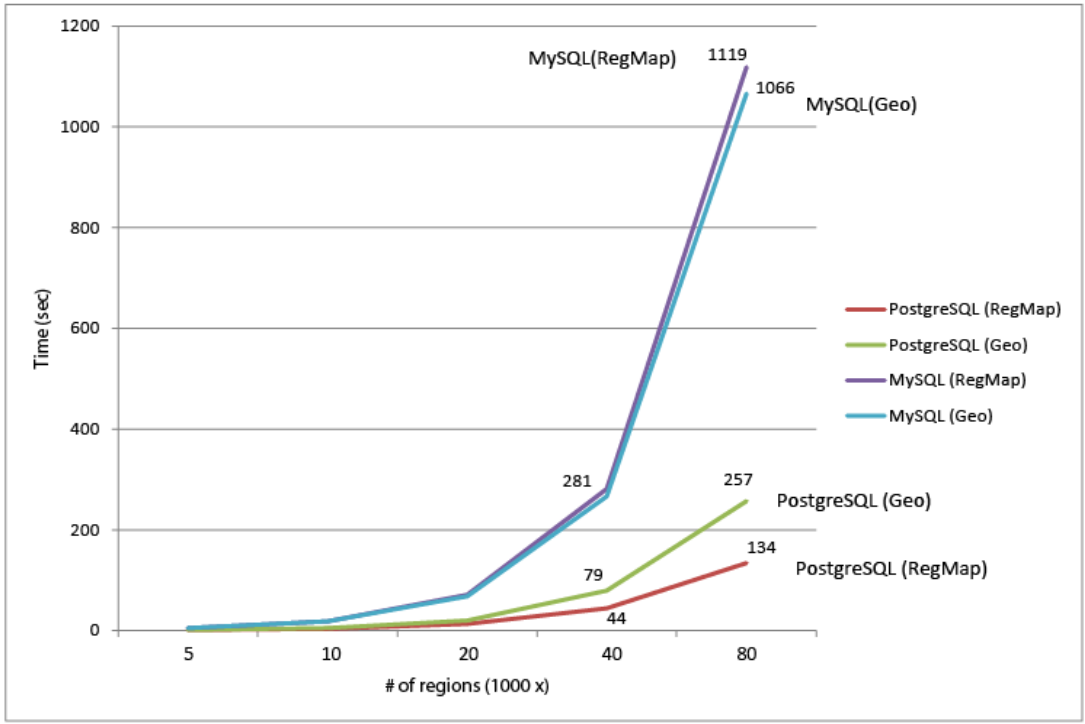

**Figure 3: Comparison of performance for identifying overlapping regions using RegMap and Geo functions.** For simplicity, InnoDB times are shown for MySQL, as the MyISAM storage engine in MySQL did not perform as well.